\documentclass[12pt]{article}
\usepackage{graphics,epsfig}
\date{}

\begin{document}

\title{IMPEDANCES AND POWER LOSSES \\ FOR AN OFF-AXIS BEAM\thanks
{Presented at the 5th European Particle Accelerator Conference,
Sitges (Barcelona), Spain, June 10-14, 1996}}

\author{Sergey~S.~Kurennoy\\
Physics Department, University of Maryland, \\
College Park, MD 20742, USA}

\maketitle

\begin{abstract} 
A method for calculating coupling impedances and power losses for 
off-axis beams is developed. It is applied to calculate impedances 
of small localized discontinuities like holes and slots, as well as 
the impedance due to a finite resistivity of chamber walls, in 
homogeneous chambers with an arbitrary shape of the chamber cross 
section. The approach requires to solve a two-dimensional electrostatic 
problem, which can be easily done numerically in the general case, while 
for some particular cases analytical solutions are obtained.
\end{abstract} 

\section{Introduction} 

The beam-chamber coupling impedances, as well as power losses due to
a finite conductivity of the chamber wall, may depend essentially on
the beam position inside the chamber. While for the power loss in a 
circular pipe this dependence is well-known \cite{AChao}, developing 
an approach working for other chamber cross sections seems to be 
worthwhile.

In the present note, we consider the problem for the vacuum chamber 
with an arbitrary but constant cross section, and calculate, for
an off-axis beam, the coupling impedance due to either resistive wall
or a small localized discontinuity, like a hole. Analytical results
are presented for circular and rectangular cross sections. 

\section{Longitudinal Impedance}

Let us consider an infinite cylindrical chamber with an arbitrary 
cross section $S$. The $z$ axis is directed along the chamber axis,
an ultrarelativistic point charge $q$ moves parallel to the axis 
with the transverse offset $\vec{a}$ from it. A small discontinuity 
(e.g., a hole) located on the chamber wall at the point ($\vec{b},z=0$), 
contributes as an inductance to the longitudinal coupling impedance 
\cite{SK&RLG,SK92}
\begin{equation}
 Z(k;\vec{a}\,) = - ikZ_0e^2_\nu(\vec{a}\,) \left (\psi - 
        \chi \right )/2 \ ,                              \label{Z} 
\end{equation}
where $Z_0 = 120 \pi$~Ohms is the impedance of free space, 
$k=\omega/c$, and $\psi$ and $\chi$ are magnetic and electric 
polarizabilities of the discontinuity . The dependence on the beam 
position, as well as on the hole position in the cross section, is via
\begin{equation}
 e_\nu(\vec{a}\,) = - \sum_s k^{-2}_s
 e_s(\vec{a}\,) \nabla _{\nu}e_s(\vec{b}\,)  \label{enorm}
\end{equation}
which is merely the normalized electrostatic field produced at the 
hole location by a filament charge displaced from the chamber axis 
by distance $\vec{a}$. It is expressed in terms 
of eigenvalues $k^2_{nm}$ and orthonormalized eigenfunctions (EFs) 
$e_{nm}(\vec{r})$ of the 2D boundary problem in $S$:
\begin{equation}
\left (\nabla ^2+ k^2_{nm}\right ) e_{nm} =
  0 \ ; \qquad e_{nm}\big\vert_{\partial S} = 0 \ . \label{boundpr}
\end{equation}
We denote $\hat{\nu}$ and $\hat{\tau}$ the outward normal and tangent 
unit vectors to the boundary $\partial S$ of the chamber cross 
section $S$, so that $\{ \hat{\nu},\hat{\tau},\hat{z}\}$ form 
a right-handed basis. One should note the normalization condition
\begin{equation}
 \oint_{\partial S}\! dl \ e_\nu  = 1 \ ,  \label{norma}
\end{equation}
where integration goes along the boundary ${\partial S}$, which 
reflects the Gauss law. It follows from the fact that 
Eq.~(\ref{enorm}) gives the boundary value of 
$\vec{e}_\nu(\vec{a}\,) \equiv -\vec{\nabla}
 \Phi(\vec{r}\,-\vec{a}\,)$, where $\Phi(\vec{r}\,-\vec{a}\,)$
is the Green function of boundary problem (\ref{boundpr}):
 $\nabla^2 \, \Phi(\vec{r}\,-\vec{a}\,) = - 
       \delta(\vec{r}\,-\vec{a}\,)$.
For the symmetric case of an on-axis beam in a circular pipe 
of radius $b$ from Eq.~(\ref{norma}) immediately follows 
$e_\nu(0)=1/(2\pi b)$.

Likewise, a finite resistivity of the chamber wall leads to the
resistive impedance per unit length of the chamber, e.g.\
 \cite{RLGetc92}, 
\begin{equation}
 Z_L(k;\vec{a}\,)/L = Z_s(k) \oint_{\partial S}\! dl \ 
   e^2_\nu(\vec{a}\,) \ ,                \label{Zres} 
\end{equation}
where the surface impedance $Z_s(k)$ is equal to $Z_0k\delta/2$ 
when skin-depth $\delta$ is smaller than the wall thickness.

Therefore, the problem of the impedance dependence on the beam 
position is reduced to evaluating $e_\nu(\vec{a}\,)$, cf.\ 
Eqs.~(\ref{Z}) and (\ref{Zres}). It can be performed analytically
for simple cross sections when the EFs are known, or numerically
in a general case, applying any 2D electrostatic code and imposing
(\ref{norma}) for normalization of a numerical solution.

\section{Beam-Position Dependence}

\subsection{Circular Chamber}

Using known eigenfunctions (e.g., \cite{Collin} or see \cite{KGS}) 
for a circular cross section of radius $b$, we sum up in 
Eq.~(\ref{enorm}) to get
\begin{equation}
e_\nu(\vec{a}\,)= \frac{1}{2\pi b} \ \frac{b^2-a^2}
{b^2-2ab \cos (\varphi_a -\varphi_h) +a^2} \, .   \label{ecirc}
\end{equation}
Here $a$ is the beam offset, $\varphi_a, \, \varphi_h$ are azimuth 
positions of the beam and hole. Result (\ref{ecirc}) coincides 
with the known distribution of the wall current, e.g.\ \cite{NasSach}. 
Figure~1 shows the beam-position dependence of the hole impedance 
(\ref{Z}).
Integrating in (\ref{Zres}) yields the well-known  
beam-position dependence for the power loss, e.g.\ \cite{AChao},
\begin{equation}
 \oint_{\partial S}\! dl \ e^2_\nu(\vec{a}\,) = 
 \frac{1}{2\pi b} \, \frac{b^2 + a^2}{b^2 - a^2} \, .  \label{e2circ}
\end{equation}

\subsection{Rectangular Chamber}

The eigenvalues and EFs for a rectangular chamber of width $w$ and 
height $h$ are well known, see in \cite{Collin} or \cite{KGS}.
Let a hole be located in the side wall at $x=w, \ y=y_h$. Then from 
Eq.~(\ref{enorm}) for the beam offset $\vec{a}=(x,y)$; ($|x| \le w/2, 
 \, |y| \le h/2$) from the axis at $(w/2,h/2)$ follows
\begin{eqnarray}
 e_\nu(\vec{a}\,) = \frac{2}{h} \left [ \, \sum_{n=0}^\infty 
  (-1)^n \sin \frac{(2n+1)\pi y_h}{h} \times \right. 
  \qquad \quad \nonumber\\
 \cos \frac{(2n+1)\pi y}{h} \; \frac{\sinh [(n+1/2)\pi (w+2x)/h]}
     {\sinh [(2n+1)\pi w/h]}  \label{erect} \\
 + \ \sum_{n=1}^\infty (-1)^n \sin \frac{2n\pi y_h}{h} 
     \sin \frac{2n\pi y}{h} \, \times \qquad \nonumber \\
 \left.  \frac{\sinh [n\pi (w+2x)/h]}
     {\sinh [2n\pi w/h]}  \right ]. \nonumber
\end{eqnarray}
Despite a rather long expression, this series is fast convergent
and convenient for evaluations, and it looks much simpler for a centered
beam, with $x=y=0$, cf.\ \cite{SK92}. Figure~2 shows that the impedance
increases significantly as the beam is displaced closer to the hole.

For integrated $e^2_\nu$ we obtain
\begin{eqnarray}
  \oint_{\partial S}\! dl \ e^2_\nu(\vec{a}\,) \ = \ 
   \frac{4}{w} \left [ \ \sum_{n=0}^\infty 
  \cos^2 \frac{(2n+1)\pi x}{w} \times \right.  \nonumber \\
   \ \frac{\sinh^2 [(n+1/2)\pi (h+2y)/w]}
      {\sinh^2 [(2n+1)\pi h/w]} \, +                \label{e2rect}\\
  \left. + \ \sum_{n=1}^\infty  \sin^2 \frac{2n\pi x}{w} 
  \frac{\sinh^2 [n\pi (h+2y)/w]} {\sinh^2 [2n\pi h/w]} \ \right ] 
                                               \nonumber \\
  + \ \left \{ x \leftrightarrow y; \ w \leftrightarrow h 
    \right \}  \ .  \nonumber
\end{eqnarray}
An example of a square pipe is illustrated in Fig.~3.

For a centered beam, i.e.\ $x=y=0$, it is reduced to 
\begin{equation}
 \oint_{\partial S}\! dl \ e^2_\nu(0) = 
 \frac{1}{w} \, \sum_{n=0}^\infty  
 \cosh^{-2} \frac{(2n+1)\pi h}{2w}  
 + \left \{w \leftrightarrow h  \right \} ,   \label{e2r0}
\end{equation}
the result obtained in \cite{RLGetc92}, which was also expressed 
in a closed form in terms of elliptic integrals \cite{RGO}.

\section{On Transverse Impedance}

The longitudinal and transverse wake functions are related by
Panofsky-Wenzel theorem
\begin{equation}
 \vec{\nabla\,} W(z,\vec{a}\,) = \frac{\partial}
   {\partial z} \vec{W}_\bot (z,\vec{a}\,) \ . \label{PW}
\end{equation}
The longitudinal wake function corresponding to the inductive 
impedance (\ref{Z}) of the hole is 
 $W(z,\vec{a}\,)=\delta'(z)F(\vec{a}\,)$, 
where $F(\vec{a}\,)=Z_0 e^2_\nu(\vec{a}\,) (\psi - \chi)/2$.
Together with Eq.~(\ref{PW}), it implies 
$\vec{W}_\bot (z,\vec{a}\,) = \delta(z) \vec{\nabla\,} F(\vec{a}\,)$,
and the monopole transverse impedance defined as the Fourier 
transform of $\vec{W}_\bot (z,\vec{a}\,)$ in $\tau=z/c$, is
\begin{equation}
  \vec{Z}^{mon}_\bot (k,\vec{a}\,) =  \frac{1}{c} \, 
 \vec{\nabla\,}  F(\vec{a}\,) =  Z_0 \frac{\psi - \chi}{2} 
   \vec{\nabla\,} e^2_\nu(\vec{a}\,) \ .               \label{Zt0}
\end{equation}
Defined in such a way $\vec{Z}^{mon}_\bot$ has dimension of Ohms,
and can be easily calculated when $e_\nu(\vec{a}\,)$ is found, e.g.\
Eqs.~(\ref{ecirc}) or (\ref{erect}). 
In an axisymmetric pipe, $\vec{Z}^{mon}_\bot = 0$, e.g.\ \cite{AChao}, 
which formally follows from the fact that $Z_{long}$ is independent 
of the beam position in such a case. However, presence of a hole 
breaks this symmetry, so that $\vec{Z}^{mon}_\bot$ does not vanish
even on the axis. For example, for a circular chamber with a hole
\begin{equation}
\vec{Z}^{mon}_\bot(k,0) = Z_0 \frac{\psi - \chi}{4\pi^2 b^3} 
   \vec{h\,} \ ,                                     \label{Zt0c0}
\end{equation}
where $\vec{h\,}$ is a unit vector from the axis toward the hole.
The presence of a second, symmetric hole (or a few of them) restores
the symmetry, and this effect disappears.  

The transverse kick obtained by a test charge $q_t$ which follows, 
at distance $z \ge 0$, the leading charge $q_s$, is 
\begin{equation}
\vec{p}_\bot (z,\vec{a}\,) = \frac{q_t q_s}{c} 
 \vec{W}_\bot (z,\vec{a}\,) = \frac{q_t q_s}{c} 
  \delta(z) \vec{\nabla\,} F(\vec{a}\,) \ .   \label{tkick}
\end{equation}
As an example, Fig.~4 shows the direction and magnitude of the 
monopole impedance and corresponding transverse kick in a circular 
pipe. For a rectangular chamber, the picture is similar. 
The result (\ref{tkick}) looks suspicious due to
$\delta(z)$, which means there is no influence on any test charge
with $z>0$, while self-influence of the source charge diverges. 
One should attribute this unphysical behavior to the approximations
used: (i) point-like discontinuity, (ii) ultrarelativistic charge,
and (iii) instant induction of effective dipoles on the hole. 
A rigorous approach, taking into account $\beta <1$ and a 
finite hole size, would lead to a more appropriate longitudinal
dependence, although calculations will be certainly complicated.
An involved direct calculation (using the method of the second paper
of Ref.~\cite{SK&RLG}, again with $\beta =1$) of the integrated 
transverse force acting on an on-axis charge passing a hole in a 
circular pipe leads to divergent sums which, however, would be 
natural to put equal to zero\footnote{Remark due to R.L.~Gluckstern}. 
Anyway, this question remains open.

The more usual dipole transverse coupling impedance in the chamber 
with a hole, e.g.\ \cite{SK92,SKrev,KGS}, reflects the influence of
a couple of opposite-charged particles with transverse offsets 
$(\vec{s},-\vec{s})$ on a test charge with offset $\vec{t}$:
\begin{equation}
  \vec{Z}^{dip}_\bot (k,\vec{s},\vec{t}\,) = -i Z_0 
 \frac{\psi - \chi}{2} \; \frac{e_\nu(\vec{s}\,)-e_\nu(-\vec{s}\,)}
 {2s} \, \vec{\nabla}e_\nu(\vec{t}\,) \ ,             \label{Zt1}
\end{equation}
where the limit $\vec{s} \to \vec{t} \to 0$ is usually assumed.
If instead one considers $\vec{t} \to 0$ while keeping 
$\vec{s}=\vec{a}$ finite, we get corrections to the 
transverse dipole impedance. For example, in a circular pipe
($\varepsilon=a/b<1$)
\begin{eqnarray}
  \vec{Z}^{dip}_\bot (k,\vec{a}) & = & -i Z_0 \frac{\psi - \chi}
{2\pi^2 b^4}\; \vec{h}\, \cos (\varphi_a -\varphi_h) \times \label{Zt1c}
 \\ &  & \ \frac{1-\varepsilon^2}{(1+\varepsilon^2\,)^2 
    -4\varepsilon^2 \cos^2 (\varphi_a -\varphi_h) } \ . \nonumber    
\end{eqnarray}
In the limit of $a \to 0$ it reproduces the known result for the 
transverse dipole impedance of the hole, the first line in 
(\ref{Zt1c}), cf. \cite{SK&RLG,SK92}. It corresponds to the 
deflecting force directed toward (or opposite to) the hole with its 
magnitude proportional to the beam offset and depending on beam 
azimuth position $\varphi_a$ as $\cos (\varphi_a -\varphi_h)$. 
Expanding in powers of $\varepsilon$ yields sextupole term and 
higher-order corrections:
$$ \cos{(\varphi_a -\varphi_h)} + \varepsilon^2
 \cos{3(\varphi_a -\varphi_h)} + O\left(\varepsilon^4\right) \ . $$
Results for rectangular pipes are obtained in a similar way from 
Eq.~(\ref{erect}) in terms of series.

\begin{figure}[b]
\centerline{\epsfig{figure=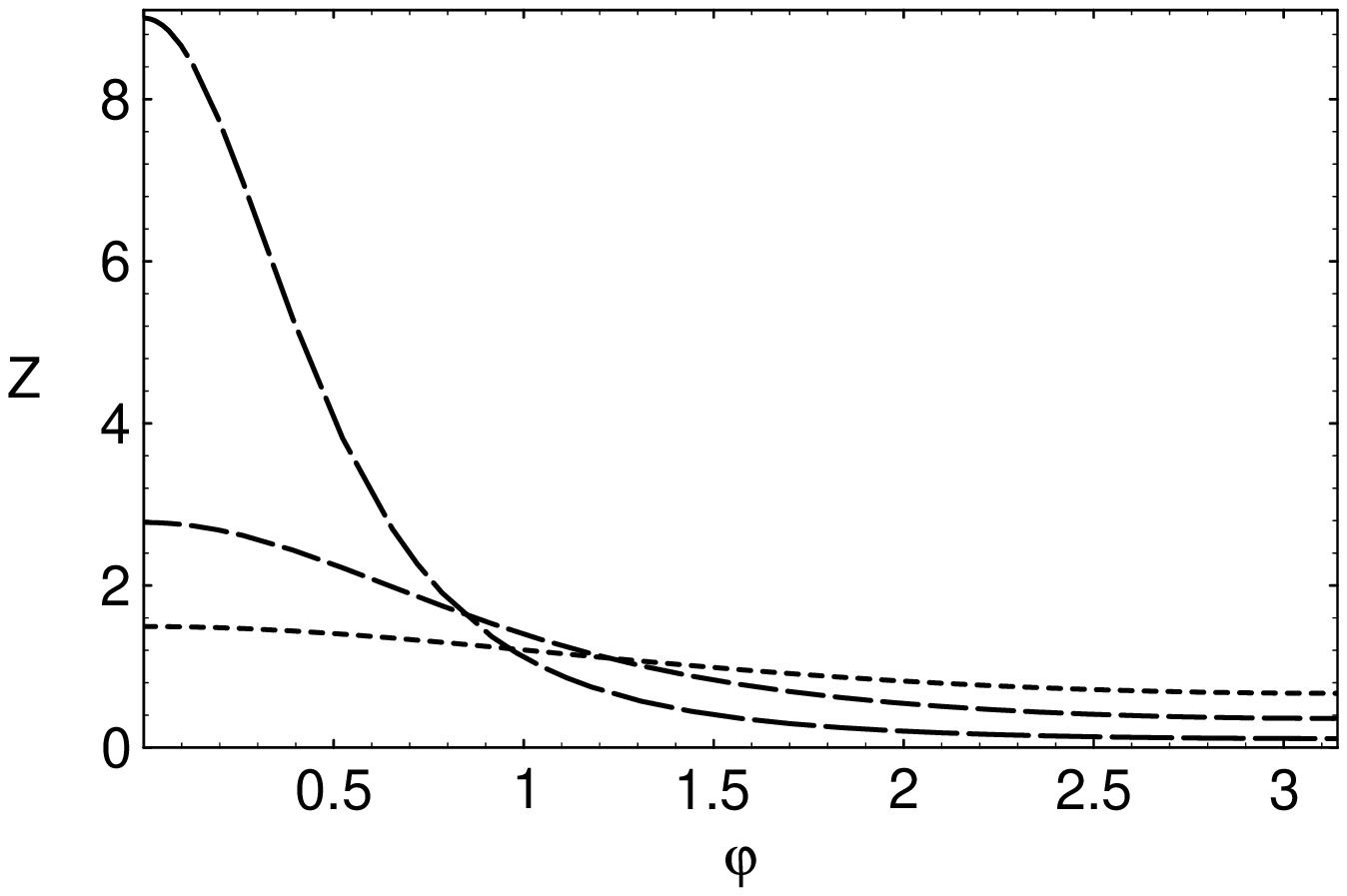,width=15cm}}
\caption{Impedance of a hole in circular pipe versus azimuth angle 
$\varphi = \varphi_a  - \varphi_h$ between beam and hole (in radians) 
for different beam offsets $a/b =$ 0.1 (short-dashed), 0.25, and 0.5 
(long-dashed). $Z=1$ corresponds to $a=0$ (on-axis beam).}
\end{figure}

\begin{figure}
\centerline{\epsfig{figure=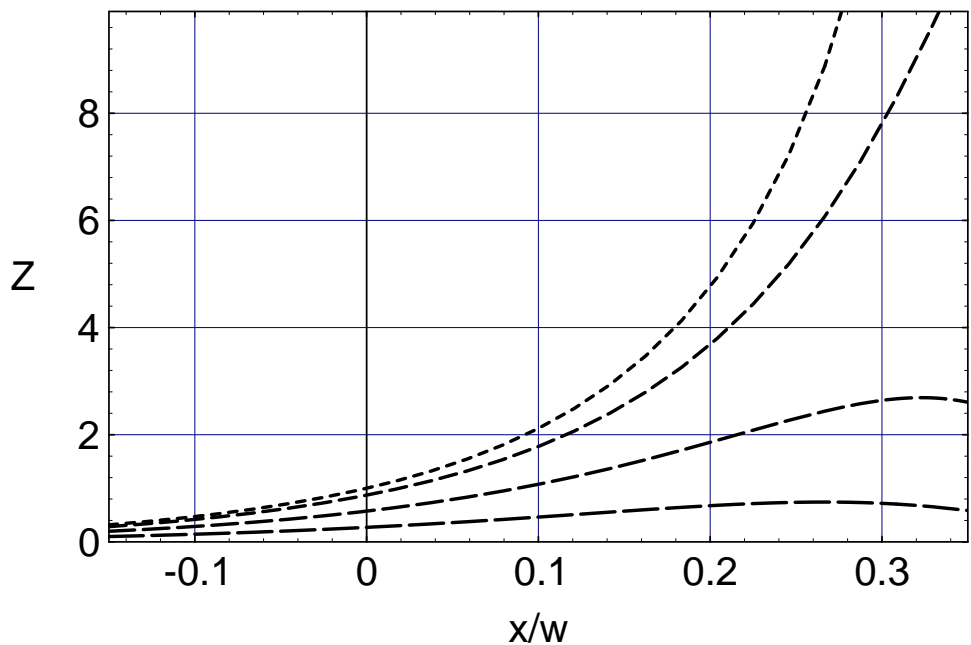,width=15cm}}
\caption{Impedance of a hole in the middle of square-pipe wall 
($y_h/h=1/2$) versus horizontal beam offset for different vertical
beam offsets $y/h =$ 0 (short-dashed), 0.1, 0.2 and 0.3 (long-dashed). 
For an on-axis beam $Z=1$.}
\end{figure}

\begin{figure}
\centerline{\epsfig{figure=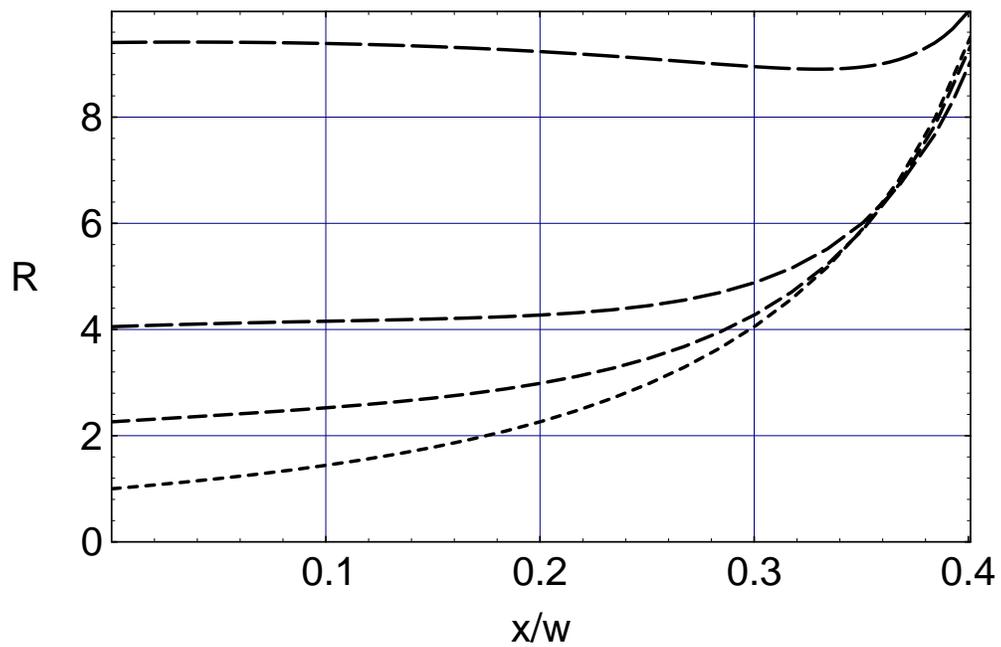,width=15cm}}
\caption{Power loss in square pipe versus horizontal beam offset
for different vertical beam offsets $y/h =$ 0 (no offset, short-dashed), 
0.2, 0.3, and 0.4 (long-dashed). $R=1$ corresponds to an on-axis beam.}
\end{figure}

\begin{figure}
\centerline{\epsfig{figure=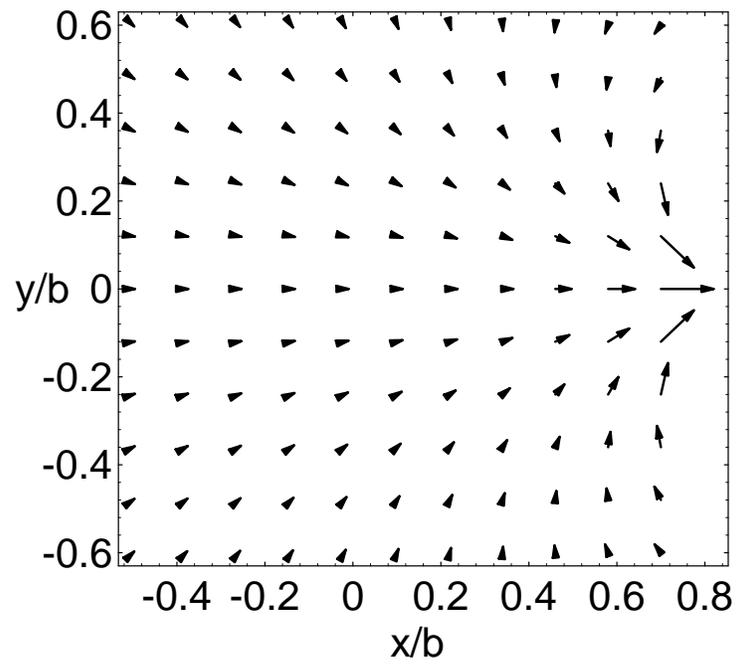,width=11cm}}
\caption{Direction and magnitude of monopole transverse impedance in 
central region of circular pipe with a hole at $\varphi_h = 0$ 
($x=b, \, y=0$) versus beam position, normalized to that magnitude 
for an on-axis beam.}
\end{figure}

\end{document}